*This article has been accepted for publication in Computers, Materials & Continua Journal.*

*Computers, Materials & Continua*

Tech Science Press

*Article*

# An Efficient Internet Traffic Classification System Using Deep Learning for IoT

Muhammad Basit Umair[1], Zeshan Iqbal[1], Muhammad Bilal[2], Tarik Adnan Almohamad[3,4], Jamel Nebhen[5], Raja Majid Mehmood[6,*]

[1]Department of Computer Science, University of Engineering and Technology Taxila, 47050, Pakistan.
[2]Division of Computer and Electronics Systems Engineering, Hankuk University of Foreign Studies, Yongin-si, Rep. of Korea.
[3]Electrical-Electronics Engineering Department, Faculty of Engineering, Karabük University, 78050, Karabük, Turkey.
[4]Department of Electronics and Communication Engineering, A'Sharqiyah University, Ibra 400, Oman
[5]Prince Sattam bin Abdulaziz University, College of Computer Engineering and Sciences, P.O. Box 151 Alkharj 11942, Saudi Arabia.
[6]Information and Communication Technology Department, School of Electrical and Computer Engineering, Xiamen University Malaysia, Sepang 43900, Malaysia.
[*]Corresponding Author: Raja Majid Mehmood. Email: rmeex07@ieee.org

**Abstract:** Internet of Things (IoT) defines a network of devices connected to the internet and sharing a massive amount of data between each other and a central location. These IoT devices are connected to a network therefore prone to attacks. Various management tasks and network operations such as security, intrusion detection, Quality-of-Service provisioning, performance monitoring, resource provisioning, and traffic engineering require traffic classification. Due to the ineffectiveness of traditional classification schemes, such as port-based and payload-based methods, researchers proposed machine learning-based traffic classification systems based on shallow neural networks. Furthermore, machine learning-based models incline to misclassify internet traffic due to improper feature selection. In this research, an efficient multilayer deep learning based classification system is presented to overcome these challenges that can classify internet traffic. To examine the performance of the proposed technique, Moore-dataset is used for training the classifier. The proposed scheme takes the pre-processed data and extracts the flow features using a deep neural network (DNN). In particular, the maximum entropy classifier is used to classify the internet traffic. The experimental results show that the proposed hybrid deep learning algorithm is effective and achieved high accuracy for internet traffic classification, i.e., 99.23%. Furthermore, the proposed algorithm achieved the highest accuracy compared to the support vector machine (SVM) based classification technique and k-nearest neighbours (KNNs) based classification technique.

## 1 Introduction

Nowadays, Internet of things (IoT) devices are connected, embedded, and producing a substantial amount of data [1], [2]. Hence, causing various network management issues for both academia and industry and opens new research areas and directions [3], [4]. Internet traffic classification (ITC) plays an important role in improving network performance, intrusion detection, QoS provisioning, and dynamic access control.

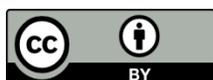

Therefore, various ITC techniques have been proposed in recent years: the port-based classification method, payload-based classification method, and machine-learning-based classification system [5].

In the port-based classification method, the port numbers are allocated by the Internet Assigned Numbers Authority (IANA) [6]. Earlier, fixed port numbers were used by most internet protocols. Due to security reasons, the majority of the protocols are using a dynamic port allocation system nowadays [7]. Hence, the dynamic port allocation limits the use of port-based traffic classification. Therefore, to overcome these restrictions, the payload-based technique was introduced. In this method, the application-layer information of packets is examined, and classification is performed by inspecting the packet payload and comparing the signature of the protocols stored in the database [8]. However, this technique is challenging to implement. For example, in the presence of encrypted traffic, the DPI cannot examine the packet payload. Moreover, from the users' perspective, DPI can cause issues related to privacy.

Machine learning-based traffic classification was proposed to overcome the challenges of payload-based and port-based methods [9], [10]–[15]. The machine learning algorithms can be classified into three categories: supervised, unsupervised, and semi-supervised. In supervised learning, a labelled dataset is used for training. In semi-supervised, both labelled and unlabeled data is used. In unsupervised learning, unlabeled data is used for classification [16], and classes are unknown. A number of machine learning algorithms such as SVM, Backpropagation, KNNs, and C4.5 [5] have been used for traffic classification [17]. These algorithms use the port number, inter-arrival time, and packet size for the classification [18]. However, there are a few challenges while using machine learning for ITC: the rapid development of new applications and new training models are required for newly emerging topologies.

In this research, A multilayer deep neural network (DNN) with a maximum entropy classifier is proposed for ITC. First, Moore's dataset is acquired and pre-processed using the OneHotEncoding technique to encode categorical variables. The pre-processed data is split into training and testing data. Next, the common flow features are identified and selected using an extra-trees classifier from pre-processed data. Finally, a feedforward DNN is trained, and the maximum entropy classifier is used at the output layer of the DNN. The proposed multilayer DNN has an accuracy of 99.23%. Furthermore, for comparison purposes, a series of experiments are conducted using SVM and KNN algorithms. The SVM and KNN algorithms achieved an accuracy of 98.90% and 98.56%, respectively. This comparison shows that multilayer DNN can be used to accurately classify network traffic as compared to shallow networks.

In summary, the major contributions of this research are as follows.

1. An efficient hybrid deep learning model is designed using a multilayer feedforward DNN, and a maximum entropy classifier for internet traffic classification. Moore's dataset is acquired and pre-processed using the OneHotEncoding technique to encode categorical variables. The pre-processed data is split into training and testing data. Features are extracted through a multilayer feedforward DNN, and a maximum entropy classifier is used to classify the internet traffic.
2. The functionalities of multilayer DNN are improved using the dropout layer. The dropout layer provides the functionality to avoid overfitting.
3. In order to eliminate the stochastic gradient descent (SGD), slow convergence, vanishing learning rate, and high variance in the parameter updates, which causes loss function to fluctuate, Adam optimizer is used for training the network.
4. A comparison of the proposed hybrid deep learning model is presented with state-of-the-art models on a benchmark dataset for internet traffic classification. Furthermore, for comparison purposes, a number of experiments are conducted using SVM and KNN algorithms.

The remainder of this paper is organized as follows. First, related works are presented in Section 2. Section 3 describes the proposed methodology for ITC. In section 4, we summarize an experimental setup. Next, the experimental results are given in section 5. Finally, the conclusion of the research and future work is discussed in section 6.

**Table 1:** A brief comparison of existing techniques

| Author | Technique used | Year of publication | Dataset | Classification Accuracy | Summary |
|---|---|---|---|---|---|
| Shi et al. [9] | Deep learning | 2018 | Moore and UNIBS | 98% | To remove the irrelevant features, the symmetric uncertainty was applied. |
| Yu et al. [10] | Semi-supervised Learning | 2018 | Campus network | Not mentioned | A DPI based method was used to label the network traffic. |
| Zhang et al. [19] | Deep Learning | 2018 | Moore-dataset | 91.21% | A DNN was designed using autoencoder and softmax model for ITC. |
| Lopez-Martin et al. [20] | RNN and CNN | 2017 | RedIRIS dataset | 0.96 | A deep learning-based model by combining CNN and RNN was employed. |
| Sun et al. [21] | Transfer learning | 2018 | Moore-dataset | 98.7% | A TrAdaBoost was to label the traffic, and Maxnet was used as the base classifier. |
| Garg et al. [22] | Deep learning | 2019 | KDD99 University dataset | Not mentioned | A deep learning method to detect the abnormal activities in SDN. |
| Ertam and Avcı [23] | Extreme learning machine | 2016 | Moore-dataset | 96.25% | To classify the network traffic using extreme learning machine. |
| K.Dias et al. [24] | Machine-learning | 2019 | Campus network | 98.88% | A real-time video classification scheme was introduced. |
| Gómezv et al. [25] | Machine-learning | 2019 | BarcelonaTech network traffic | --- | To deal with the class imbalance problem, a new approach was adopted |
| Lotfollahi et al. [26] | CNN and encoder | 2020 | ISCX VPN-nonVPN | --- | A novel approach has been provided using CNN and Stacked auto-encoder |
| Cao et al. [27] | SVM | 2020 | Moore Dataset | 91.96% | A wrapper-based hybrid feature selection method to select important features. |

**2 Related Work**

In the last decade, numerous research have been conducted for ITC, such as port-based, payload-based, and machine-learning-based approaches. Recently, machine-learning-based techniques have gained more interest due to their high performance in ITC [5]. Many algorithms are applied to the network traffic classification, but these algorithms are very different in nature. Most of the researchers focused on feature and machine learning algorithm selection. Both the feature and algorithm selection have great importance in improving the classifier performance. A summary of the relevant and state of the art classification methods, the techniques used, their dataset usage, accuracy, and associated algorithms are presented in Tab. 1.

Zhang et al. [19] have proposed a technique for application classification based on deep learning and software-defined networks (SDN) architecture. A hybrid deep learning model for network application classification was proposed by merging the stacked auto-encoder and the softmax regression. A stacked auto-encoder was used for flow features extraction, and the softmax regression layer was used for classifying the network application. The author achieved an average accuracy of 91.21% using five hidden layers and ten hidden nodes.

Lopez-Martin et al. [20] suggested a deep learning model which combines a convolutional neural network (CNN) and long short-term memory (LSTM) to identify the network traffic. For every flow, different features were extracted from the packet's header, and a time series feature vector was built. More than 25000 features and 100 services were used to train a model. On the best trained model, this statistical approach achieved 0.9632 and 0.9574 of accuracy and F1-score, respectively. Sun et al. [21] introduced a TrAdaBoost system that utilized the labeled traffic data collected from different sources for the classification of network traffic. A base classifier, the maximum entropy model (Maxent), was used to implement a source of knowledge from source data to target data. An accuracy of 98.7% was obtained using the TrAdaBoost model.

To identify suspicious flows in SDN based networks, Garg et al. [22] presented a anomaly detection technique using deep learning for social media domains. Two refined algorithms were used to satisfy the QoS requirements of the SDN. However, the dataset used in this research is not flow-based SDN. A kernel-based Extreme Learning Machine (ELM) approach [23] was applied to Moore's dataset to classify the internet traffic. To select the best features, a genetic algorithm was used. There were 12 attributes in the flow features. They obtained 96.25% accuracy using the wavelet activation function.

A video classification method [24], based on the Naïve Bayes classifiers, was implemented for real-time network traffic classification. There were three classes in the proposed classification scheme: one file download service and two video services. The author stated that the accuracy of 98.88% was achieved in real-time scenarios. This method cannot be implemented in the case of encrypted traffic, and it is computationally expensive. To handle the class imbalance problems, Gómezv et al. [25] used a base estimator to form a baseline.

Lotfollahi et al. [26] introduced a method named "deep packet" using one-dimensional convolutional neural network (1-D CNN) and stacked auto-encoder. This model was trained to classify the encrypted network traffic. This model achieved 98% and 94% of recall for application identification and traffic categorization, respectively. Cao et al. [27] proposed a network traffic classification technique based on SVM. To prevent overfitting, a wrapper-based feature selection algorithm is used to select important features. The authors have achieved good classification accuracy of 91.96% for multi-class classification, which was 4.2% higher than the original SVM. Kordestani et al. [28] surveyed a list of methods used for fault diagnosis and prognosis. For encrypted traffic classification, Aceto et al. [29] designed a method using a deep learning algorithm. In this method, first, they discussed deep learning architecture for encrypted traffic classification. They used 1-D CNN, 2-D CNN, LSTM, stacked autoencoder (SAE), and multilayer perceptron algorithms that can identify up to 49 mobile apps.

In [30], Aceto et al. proposed a novel, sophisticated multimodal deep learning framework for multi-class mobile traffic classification based on network packets. They take two different time series and payload-based features. In the multimodal deep learning architectures, the obtained accuracy 79.6% for FB/FBM class, and 89.49%,89.14% for android and IOS, respectively. For feature engineering, Swarna et al. [31] used a hybrid technique using Principal Component Analysis (PCA) and Grey Wolf Optimization (GWO) for intrusion detection in Internet of Medical Things and achieved good results. An Efficient Feature Optimization Approach (EFOA) [9] was proposed for optimal feature selection and optimization. A series of experiments were performed on the University of Cambridge dataset [32]. First, the correlation of original flow statistics was evaluated, and irrelevant features were removed using symmetric uncertainty (SU). Next, to get the robust features, related features were passed to the feature generation model. The feature generation model was based on a deep belief network, implemented using unsupervised learning. Finally, redundant features were removed using weighted symmetric uncertainty (WSU). However, the proposed model is computationally expensive.

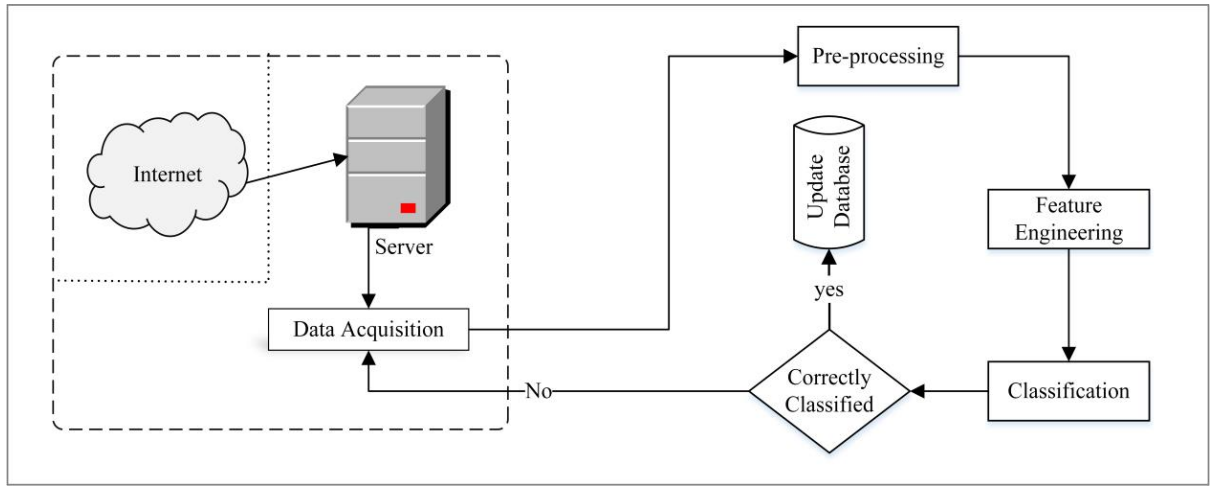

**Figure 1:** Overview of the proposed system methodology

On the other hand, in the semi-supervised based classification methods, Yu et al. [10] combined the DPI and semi-supervised learning for multi-classifier in SDN. The combined arrangement was used to classify applications into different categories. A DPI-based technique was used to maintain a traffic database, and a partially labeled dataset was formed and stored. However, this method cannot classify traffic if the data is encrypted. Reddy et al. [33] discussed two methods PCA and GWO, for dimensionality reduction using different machine learning algorithms.

Although numerous research efforts have been invested for ITC, there are still prominent issues in the existing literature that need to be addressed, including incapability of classifying encrypted traffic, less accuracy, and computationally expensive algorithms. The other problem with the aforementioned techniques is that the detailed number of classes is missing. An efficient method is proposed to overcome these issues that consist of a deep neural network and a softmax classifier. Compared with other networks, the proposed method not only efficiently classifies the internet traffic but also it is computationally less expensive. The performance and obtained results of the proposed system are compared with recent and state-of-the-art ITC schemes. Furthermore, the proposed study provides a complete roadmap for ITC.

## 3 Proposed Methodology

An abstract view of the methodology used throughout this work is shown in Fig. 1. The proposed methodology consisting of four key steps: data acquisition, pre-processing, feature engineering, and classification. The acquired data is processed using LabelEncoder and OneHotEncoder to make the classification of traffic useful. Thus, the proposed scheme is aware of classifying internet traffic into different classes. In the case of real-time traffic, the proposed technique constantly acquires data flows from the internet in the form of flow statistics.

### 3.1 Data Acquisition (Dataset)

This work employed the Moore-dataset, which contains a real-world traffic traces dataset with 12

continuous features. These traffic traces have been collected from the University of Cambridge, publicly available for research purposes [32]. The Cambridge University traffic traces are released by the computer lab on Genome Campus. The major advantage of these traffic traces is that it provides a base to get more accurate results and is extensively acceptable for assessment and comparison of traffic classification approaches. Therefore, it can be used as a benchmark to assess the performance of the ITC algorithm. These Cambridge traces are being used by most researchers for ITC research [19], [21], [34], [35], [36].

There are 324,277 data flows samples in the dataset. The imperative advantage of selected Moore-dataset is different flows for different classes. The acquired dataset consists of TCP traffic flows and hand-verified class labels (ground truth). Li et al. [34] used these traffic traces for research purposes. All the flows are classified into seven application classes, i.e. Mail, P2P, WWW, Chat, Bulk, Database, and Interactive, and other classes are excluded due to the minority number of samples in the dataset.

### 3.2 Data Pre-processing

Afterwards, acquisition of the dataset, pre-processing of the data is an essential step. In supervised machine learning, it plays a key role in improving the performance of classifiers. Data pre-processing helps in reducing the computational cost of any system. In the Moore-dataset [32], LabelEncoder converts categorical values into numeric values using the scikit-learn library. In this proposed methodology, there are seven distinct classes, and LabelEncoder encodes the labels with unique numeric values among 0 and (n-1), where n represents the number of distinct classes. OneHotEncoder takes each column in categorical data, which are label encoded and converted into binary columns. These columns are replaced by 0s and 1s, for each category correspondence to which column has been replaced. Min-max normalization is performed to scale the dataset values between the range 0 and 1 by the given Eq. (1).

$$x' = \frac{x - x_{min}}{x_{max} - x_{min}} \quad (1)$$

Where $x'$ represents the normalized value, $x_{max}$ is the maximum value, and $x_{min}$ is the minimum value in the dataset.

### 3.3 Feature Engineering

Machine learning models may face many problems due to sparse features. This feature selection technique is used to overcome the sparse features and select the most essential features to improve the performance of machine learning models. For the identification of important features from the pre-processed dataset, the extra-trees classifier is used. The feature importance function calculates the information gain of all features and returns the high information gain features. This feature extraction system shows better performance as compared to previous approaches related to ITC [21], [23]. To make the proposed methodology computationally efficient, flow-based features are selected. Feature sets extracted for network traffic classification are shown in Tab. 2. There are 12 features in Feature_set F (2) and correspondence class.

$$F = \{f1, f2, f3, ..., f12\} \quad (2)$$

Where $f1$ represents the server port number and $f2$ represents the client port number, and detail of other features is given in Tab. 2. Finally, all classes and their correspondence applications are given in Tab. 3.

**Table 2:** The information of given extracted features

| Feature | Description |
|---|---|
| f1 | Server_port |
| f2 | Client_ port |
| f3 | Actual_data_packets (c to s) |
| f4 | Pushed_data_packets (c to s) |
| f5 | Pushed_data_packets (s to c) |
| f6 | Minimum_segment_size (c to s) |
| f7 | Average_segment_size (c to s) |
| f8 | Initial_window_bytes (c to s) |
| f9 | Initial_window_bytes (s to c) |
| f10 | RTT_samples (c to s) |
| f11 | Median_data_packets (c to s) |
| f12 | Variance_bytes_packet (s to c) |
| CLASS | WWW, P2P, MAIL, INTERACTIVE, BULK, SERVICES, DATABASE |

**Table 3:** List of classes and correspondence applications

| Class | Applications | Percentage (%) |
|---|---|---|
| WWW | www | 84.077 |
| P2P | BitTorrent, GnuTella | 8.571 |
| BULK | ftp-control, ftp-pasv, ftp-attack | 2.058 |
| INTERACTIVE | telnet | 0.124 |
| MAIL | IMAP, pop2, SMTP | 1.530 |
| DATABASE | Sqlnet, oracle, ingress | 3.531 |
| CHAT | Yahoo IM, Jabber, MSN Messenger | 0.025 |

*3.4 Classification*

In this work, a hybrid deep learning system that consists of a DNN and maximum entropy classifier, also known as the softmax regression model, is employed and trained to classify network traffic. A DNN is a feedforward artificial neural network with backpropagation using multiple hidden layers. A DNN comprises an input layer, multiple hidden layers, and an output layer. An overview of the proposed DNN is depicted in Fig. 2. The input layer consists of twelve nodes. The DNN is given by Eq. (3), where $x_i$ is the input, $w_i$ is the weight, and $w_0$ is the bias of any neuron. The applied activation function is given in Eq. (4), where $y_{out}$ is the observed output. The objective of the implemented DNN is to reduce the error between the input and output layer. The loss function can be calculated by the Eq. (5). A nonlinear activation rectified linear unit (ReLU) is applied on the input layer. The ReLU is given by Eq. (6). ReLU helps in performance-boosting and accelerating the training method. The main advantage of using ReLU is the removal of exploding gradient problem.

$$y_{net} = w_i x_i + w_0 \tag{3}$$

$$y_{out} = f(y_{net}) = \frac{1}{1 + e^{-y_{net}}} \tag{4}$$

$$L = \frac{1}{2} \sum_{i=1}^{N} \| y_{net} - y_{out} \|^2 \tag{5}$$

$$\text{Re}\, LU(x) = \max(0, z) \tag{6}$$

There are several hidden layers, each consisting of sixteen neurons. Each hidden layer is connected with a nonlinear sigmoid activation function, given by Eq. (7). The ITC problem is a multi-class classification problem, so a maximum entropy classifier is used at the output layer. The maximum entropy classifier belongs to supervised learning, and is usually used with other classifiers. Mathematically, the maximum entropy classifier is given by Eq. (8).

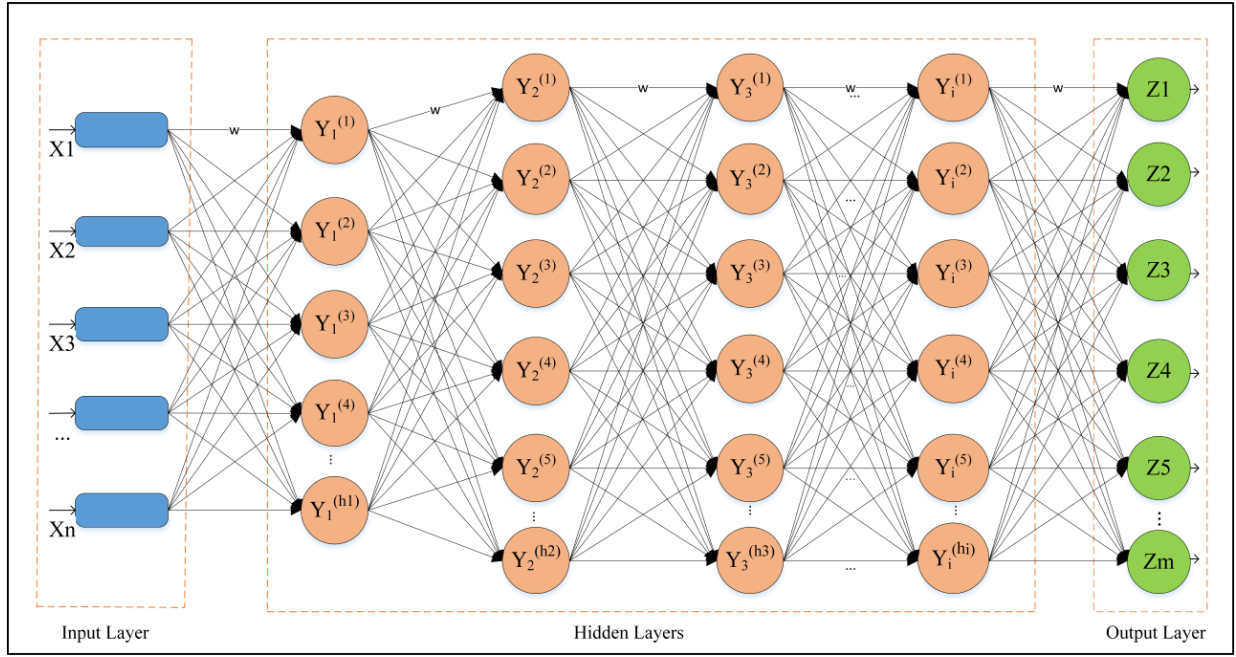

**Figure 2:** Proposed deep learning model.

$$sigmoid(x) = \frac{1}{1+e^z} \tag{7}$$

$$\sigma(u)_j = \frac{e^{u_j}}{\sum_{m=1}^{M} e^{u_m}} \tag{8}$$

Where $u$ is the M-dimensional vector, $u_m$ and $u_j$ are the elements of the M-dimensional vector. The $\sigma(u)$ represents M-dimensional after mapping $m$ and $j$. The $j, m$ are subscripts and values of $j, m = 1, 2, 3 \ldots M$.

## 4 Parameters Optimization

The performance of a deep learning model varies by the selection of hyperparameters. The grid search parameter technique is used to find out the best deep learning model in this method. To find out the best hyperparameters, the possible combination of every outcome is evaluated. The list of hyperparameters used in this study is given in Tab. 4. In this method, we verify the performance of the proposed method by performing k-fold cross-validation and the train-test split, to find out the optimal hyperparameters. In this proposed method, the batch size is 1000.

**Table 4:** Hyperparameters used for the training of DNN.

| Parameter | Value |
| --- | --- |
| Activation function | ReLU, sigmoid, Softmax |
| Loss function | Categorical cross entropy |
| Optimizer | Adam |
| Learning rate | 0.01 |
| Batch size | 1000 |
| Cross-Validation | 10 |
| Number of epochs | 500 |
| Dropout rate | 0.2 |

The categorical cross-entropy is used as a loss function. Cross-entropy achieved the best performance in multi-class classification. Moreover, the cross-entropy function has faster convergence and low complexity during the iterative optimization process. The simulation environment of the proposed methodology consists of 500 epochs. To avoid overfitting, we used dropout rate (0.2) after every two hidden layers. The dropout layer removes neurons randomly. In order to eliminate the stochastic gradient descent (SGD), vanishing learning rate, slow convergence, and high variance in the parameter updates, which causes loss function to fluctuate, Adam optimizer is used. The learning rate of 0.01 was chosen that performs best in the case of both training and testing sets. Adam is computationally efficient and an adaptive learning rate optimization algorithm [37]. Further, we used a batch normalization layer that speeds up the deep neural network training and thus relieved the gradient dissipation.

## 5 Results and Discussion

The simulation experiments are performed using an Intel Core i5 CPU @ 1.80GHz, 64-bit OS, x64-based processor on Windows 10 platform. We used Python (version 3.6.5) as a programming language, with scikit-learn library, to perform the experiments [38]. The scikit-learn is an open-source machine learning library. Scikit-learn module is used with NumPy for scientific operations [39]. The dataset is divided as a 7:3 train-test split. The division of the dataset into 70% of the training set and 30% of the test set is randomized.

### 5.1 Evaluation Metrics

There are four evaluation metrics in the proposed methodology. The explanation of each metric is given in Tab. 5.

**Table 5:** The evaluation metrics in our proposed methodology

| Metric | Definition | Mathematical Expression |
|---|---|---|
| Accuracy | A number of correctly classified packets in the overall dataset | $Acc = \frac{TP+TN}{TP+FN+FP+FN}$ |
| Precision | The ratio of true positive to entirely positive results | $Percision = \frac{TP}{TP+FP}$ |
| Recall | True positive rate | $Recall = \frac{TP}{TP+FN}$ |
| F1-score | The measure of the test accuracy | $F1-score = 2 * \frac{Percision*Recall}{Percision+Recall}$ |

### 5.2 Experimental Results

The experiments are performed to assess the performance of the DNN. From the data mining perspective, ITC is a multi-class classification problem. To find the best deep learning model, it is necessary to identify the number of hidden layers in a DNN. The DNN deals with a high number of hidden layers, and it has a stronger classification capability to achieve higher classification results. Also, the learning rate plays an important role in maintaining convergence. The experiments are performed to determine the number of hidden layers of the network.

Tab. 6 presents the results of the relationship between the number of hidden layers and DNN performance for all four evaluation metrics. It is observed from Tab. 6 that the highest accuracy of 99.23% is achieved by using the seven hidden layers. However, by increasing the number of hidden layers, the performance of the network doesn't improve; therefore, seven hidden layers are chosen for DNN used in this study.

**Table 6:** The performance of the proposed deep learning framework

| No | Hidden layers | Accuracy (%) | Precision (%) | Recall (%) | F1-score (%) |
|----|---------------|--------------|---------------|------------|--------------|
| 1  | 3             | 94.66        | 92.49         | 94.66      | 93.01        |
| 2  | 4             | 99.08        | 99.00         | 99.08      | 99.03        |
| 3  | 5             | 99.13        | 98.90         | 99.13      | 99.02        |
| 4  | 6             | 99.11        | 98.89         | 99.11      | 99.00        |
| 5  | 7             | 99.23        | 99.15         | 99.23      | 99.18        |
| 6  | 8             | 99.15        | 99.08         | 99.15      | 99.10        |
| 7  | 9             | 99.13        | 99.11         | 99.13      | 99.12        |
| 8  | 10            | 99.10        | 99.11         | 99.10      | 99.10        |

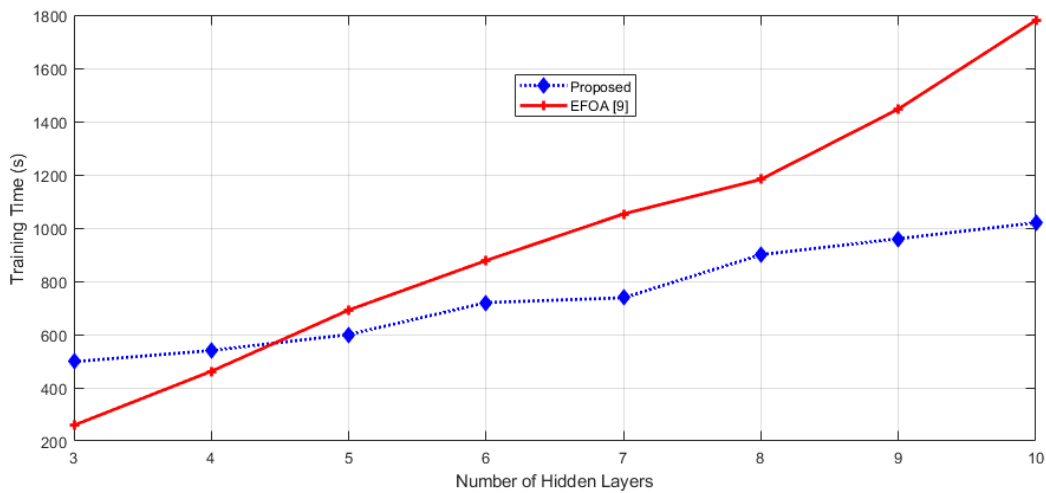

**Figure 3:** Comparison of computational time with state-of-the-art

In Fig. 3, the computational cost for the proposed algorithm is depicted. It is evident from the graph that by increasing the number of hidden layers, the proposed algorithm is computationally less expensive as compared to EFOA.

**Table 7:** Results achieved using DNN for each class

| Class       | Accuracy (%) | Precision (%) | Recall (%) | F1-Score (%) |
|-------------|--------------|---------------|------------|--------------|
| WWW         | 99.00        | 100.0         | 99.00      | 100.0        |
| P2P         | 99.00        | 100.0         | 99.00      | 99.00        |
| MAIL        | 84.00        | 85.00         | 84.00      | 84.00        |
| INTERACTIVE | 89.00        | 57.00         | 89.00      | 69.00        |
| DATABASE    | 100.0        | 100.0         | 100.0      | 100.0        |
| CHAT        | 69.00        | 50.00         | 69.00      | 56.00        |
| BULK        | 92.00        | 84.00         | 92.00      | 88.00        |

Tab. 7 shows the results achieved using the proposed DNN for each class. Four evaluation parameters are calculated to verify the performance of the network. The detail of each class is given in Tab. 7, and corresponding parameters are given in Tab. 5. For Database, WWW, and P2P classes, the accuracy, precision, recall, and F1-score are high because there is a greater number of samples in the dataset. On the other hand, each metric has a low score for the Chat class due to the lowest number of samples. The graphical representation of accuracy, precision, recall, and F1-score for each class is depicted in Fig. 4.

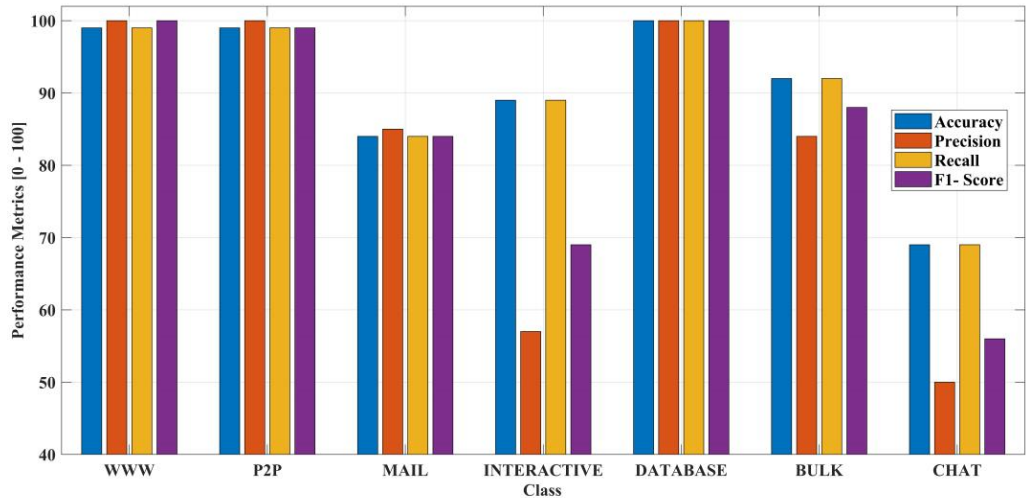

**Figure 4:** Performance metrics on different applications using DNN

## 5.3 Performance Evaluation

In k fold cross-validation, the dataset is divided into k-disjoint equal subset in which k-1 samples are used for training, and the remaining samples are used as a testing set.

**Table 8:** Results achieved using DNN for each class using k=10 folds

| Class | Accuracy (%) | Precision (%) | Recall (%) | F1-Score (%) |
| --- | --- | --- | --- | --- |
| BULK | 68.00 | 99.00 | 68.00 | 81.00 |
| DATABASE | 100.0 | 100.0 | 100.0 | 100.0 |
| INTERACTIVE | 100.0 | 98.00 | 100.0 | 99.00 |
| MAIL | 99.00 | 59.00 | 99.00 | 74.00 |
| P2P | 100.0 | 100.0 | 100.0 | 100.0 |
| WWW | 100.0 | 100.0 | 100.0 | 100.0 |

Tab. 8 presents the accuracy, precision, recall, and F1-score for different classes using k=10 fold and the sample size=2000. Finally, Tab. 9 summarizes the average accuracy, precision, recall, and F1-score for each class. It is evident from Tab. 9 that the performance of the proposed DNN is exceptional for each evaluation metric accuracy (99.15%), precision (99.29%), recall (99.15%), and F1-score (99.12%) using k-folds cross-validation for the classification of network traffic.

**Table 9:** Results achieved using DNN for using k=10 folds

| Model | Accuracy (%) | Precision (%) | Recall (%) | F1-Score (%) |
| --- | --- | --- | --- | --- |
| DNN | 99.15 | 99.29 | 99.15 | 99.12 |

Further experiments are conducted to evaluate the performance of the proposed DNN with two shallow networks: SVM and KNN. Tab. 10 summarizes the comparison results. For each evaluation metric, the proposed network achieved the highest accuracy (99.23%) in comparison to SVM (98.56%) and KNN (98.90%).

**Table 10:** Accuracy using DNN, KNN, and SVM

| Classifier | Accuracy (%) | Precision (%) | Recall (%) | F1-score (%) |
|---|---|---|---|---|
| SVM | 98.56 | 98.06 | 98.56 | 98.25 |
| KNN | 98.90 | 98.41 | 98.90 | 98.45 |
| DNN | 99.23 | 99.15 | 99.23 | 99.18 |

Tab. 11 compares the results of the proposed traffic classification approach with the different state-of-the-art research studies [9], [19], [21], [23], [27], [36], [40], [41] for ITC. It can be observed that using DNN architecture, the highest accuracy of 99.23% is achieved using the Moore dataset due to its strong classification capability.

**Table 11:** Performance comparison of proposed traffic classification scheme results with state-of-the-art.

| Study | Year of Publication | Accuracy (%) | Precision (%) | Recall (%) | F1-score (%) |
|---|---|---|---|---|---|
| [9] | 2018 | 98.0 | - | - | - |
| [19] | 2018 | 91.21 | | | 90 |
| [21] | 2018 | 98.7 | - | - | - |
| [23] | 2017 | 96.2 | - | - | - |
| [27] | 2020 | 91.96 | - | - | - |
| [36] | 2020 | 96.2 | - | - | - |
| [40] | 2021 | 94.2 | - | - | - |
| [41] | 2021 | 90 | - | - | - |
| Proposed DNN | | 99.23 | 99.15 | 99.23 | 99.18 |

## 6 Conclusion

In the research, a novel multilayer DNN is proposed for ITC. The real-world traffic, Moore-dataset, is pre-processed, and important features are selected using an extra-trees classifier. A DNN network is developed with seven hidden layers, and a maximum entropy classifier is used at the output layer to classify traffic into different classes. Four performance evaluation metrics: accuracy, precision, recall, and F1-score are calculated to evaluate the performance of the proposed methodology. Experimental results show that the proposed network has achieved the highest accuracy of 99.23% as compared to 98.56% using the SVM classifier and 98.90% using the KNN classifier. Thus, the proposed DNN can be used to accurately classify real-world internet traffic.

In machine learning algorithms, a dataset is required for the training and testing of the classifier. To achieve more accurate results in ITC, there is a need to start activities for the deployment of training datasets for artificial intelligence in computer networks.

**Funding Statement:** This work has supported by the Xiamen University Malaysia Research Fund (XMUMRF) (Grant No: XMUMRF/2019-C3/IECE/0007).

**Conflicts of Interest:** The authors declare that they have no conflicts of interest to report regarding the present study.